\documentclass[aps,pra,twocolumn,groupedaddress,showpacs]{revtex4}
\usepackage{epsfig}
\usepackage{amsmath}
\usepackage{amssymb}
\usepackage{graphicx}
\usepackage{dcolumn}
\usepackage{bm}
\begin{document}

\title{
Coherent pulse position modulation quantum cipher supported by secret key
}
\author{Masaki Sohma}
\affiliation{
Research Center for Quantum Information Science,  
Tamagawa University\\
6-1-1, Tamagawa-gakuen, Machida, Tokyo, 194-8610, JAPAN
}
\email{sohma@eng.tamagawa.ac.jp
}
\author{Osamu Hirota}
\email{hirota@lab.tamagawa.ac.jp
}
\affiliation{
Research Center for Quantum Information Science,  
Tamagawa University\\
6-1-1, Tamagawa-gakuen, Machida, Tokyo, 194-8610, JAPAN
}


\date{\today}
\begin{abstract}
A quantum cipher supported by a secret key so called keyed communication in quantum noise (KCQ)  is very attractive in implementing high speed key generation and secure data transmission.
However, Yuen-2000 as a basic model of KCQ  has a difficulty to ensure the quantitative security evaluation because all physical parameter for the cipher system is finite.
Recently, an outline of a generalized scheme so called coherent pulse position modulation(CPPM) to show the rigorous security analysis is given, where the parameters are allowed to be asymptotical.
This may open a new way for the quantum key distribution with coherent states of considerable energy and high speed.

In this paper, we clarify a generation method of CPPM quantum signal by using a theory of 
unitary operator and symplectic transformation, and show an asymptotic property of security  and its numerical examples.

\end{abstract}

\maketitle

\section{Introduction}
An application of quantum phenomena to securing optical network has received much attention. 
In this case, instead of mathematical encryption, a guarantee of security by a physical principle is expected.
So far the quantum key distribution(QKD) based on very weak optical signals has been developed   and demonstrated in many institutions. 
However, these have inherent difficulty such as quantum implementations and 
very low bit rates compared to current data transmission rates.

In order to cope with such a drawback, in 2000, a new  quantum cipher was proposed [1]. It is a kind of stream cipher with randomization by  quantum noise generated in measurement 
of coherent state from the conventional laser diode.
The  scheme is called Yuen-2000 protocol(Y-00) or $\alpha\eta$ scheme[2,3] which consists of  large number of basis to transmit the information bit and  pseudo random number generator(PRNG) for the selection of the basis. A coherent state as the ciphertext which is transmitted from the optical transmitter(Alice) is determined by the input data and the running key $K$ from the output sequence of PRNG with a secret key $K_s$. The legitimate receiver(Bob) has the same PRNG, 
so he can receive  the correct ciphertext under the small error, and simultaneously demodulate the information bit. 
The attacker (Eve), who does not know the key, has to try to discriminate all possible coherent state signals. Since the signal distance among coherent state signals are designed as very small, Eve's receiver suffers serious error to get the ciphertext.
Such a difference of the accuracy of the ciphertext for Bob and Eve brings preferable security 
which cannot be obtained in any conventional cipher. 
Unfortunately, it is very difficult to clarify the quantitative security evaluation for this type of cipher, because all physical parameter for the cipher system is finite.
So far, complexity theory approach [4] and information theoretic approach [5] have been tried,
but still there is no rigorous theoretical treatment.

Recently, Yuen has pointed out that it is possible to show the rigorous security analysis 
when the parameters are allowed to be asymptotical, and showed a sketch of the properties 
using a model of coherent pulse position modulation (CPPM) [6].
This may open a new way for the quantum key distribution with coherent states of considerable energy and high speed.

In this paper, we clarify a generation method of CPPM quantum signal by using a theory of 
unitary operator and symplectic transformation, and show a security property and 
its numerical examples. In the section II, the back ground for the information theoretic security and the Shannon limit are explained. In the section III and IV, we describe a theory of CPPM. In the section V and VI, 
we discuss on the security and implementation problem.

\section{Back ground of information theoretic security}
In the conventional cipher, the ciphertext $Y$ is determined by the information bit $X$ and running key $K$.
This is called non random cipher. However, one can introduce more general cipher system so called random cipher by noise such that the 
ciphertext is defined as follows:
\begin{equation}
Y_i=f(X_i,{K_i}, r_i)
\end{equation}
where $r_i$ is noise. Such a random cipher by noise may provide a new property in the security.
In Shannon theory for the symmetric key cipher, we have the following theorem.\\
\\
{\bf Theorem 1}(Shannon, 1949 [7])\\
The information theoretic security against ciphertext only attack on data has the  limit
\begin{equation}
H(X|Y) \le H(K_s).
\end{equation}

This is called Shannon limit for the symmetric key cipher. To be beyond the Shannon limit is essential for fresh key generation 
by communication or information theoretic security against known plaintext attack in the symmetric key cipher.
In the context of random cipher, one can exceed this limit. 
It is known that the necessary condition for exceeding the limit is $Y^E \ne Y^B$ [6,8,9].
That is, the ciphertexts for Bob and Eve are different. 
Still the necessary and sufficient condition is not clear, but if the following relation holds, one can say that 
the cipher exceeds the Shannon limit
\begin{equation}
H(X|Y^E,K_s) > H(X|Y^B, K_s) =0.
\end{equation}
This means that Eve cannot pin down the information bit even if she gets a secret key 
after her measurement of the ciphertext while Bob can do it.
In the following sections, we will clarify that CPPM has indeed such a property.

\section{Coherent pulse position modulation cryptosystem}
The coherent pulse position modulation (CPPM) cryptosystem has been proposed as a quantum cipher permitting asymptotical system parameters [1,6].

Alice encodes her classical messages by the block encoding 
where $n$-bit block $j$ ($j=1,....,N=2^n$) corresponds to
the pulse position modulation (PPM) quantum signals with $N$ slots, 
\begin{equation}
|\Phi_j\rangle=|0\rangle_1\otimes \cdots \otimes |\alpha_0\rangle_j\otimes\cdots\otimes |0\rangle_N.
\end{equation}
In addition, Alice apply the unitary operator $U_{K_i}$ to PPM quantum signals $|\Phi_j\rangle$, where 
 the unitary operator $U_{K_i}$ is randomly chosen
via running key $K_i$ generated by using PRNG  on a secret key $K_s$.
Thus, Alice gets CPPM quantum signal states,
\begin{equation}
|\Psi_{j}(K_i)\rangle=U_{K_i}|\Phi_j\rangle=|\alpha_{1j}(K_i)\rangle_1\otimes\cdots\otimes|\alpha_{Nj}(K_i)\rangle_N,
\end{equation} 
which are sent to Bob.
Let us assume an ideal channel.
Since the secret key $K_s$, PRNG  and the map $K_i \to U_{K_i}$ are shared 
by Alice and Bob, Bob can apply the unitary operator $U_{K_i}^\dagger$ to the 
received CPPM quantum signal $|\Psi_{j}(K_i)\rangle$ and obtain the 
PPM quantum signal $|\Phi_{j}\rangle$.
        Bob decodes the message by the direct detection for $|\Phi_j\rangle$, 
which is known to be a suboptimal detection for PPM signals [10].
        Then Bob's block error rate is given by 
\begin{equation}\label{Pedir}
P_e^{dir}=(1-\frac{1}{N})e^{-|\alpha_0|^2} < e^{-|\alpha_0|^2} .
\end{equation}
        Here $e^{-|\alpha_0|^2} \approx 0$ holds for enough large 
signal energy $S=|\alpha_0|^2$. 
        In contrast, Eve does not know the secret key $K_s$ and hence
she must detect  CPPM quantum signals  directly.
This makes Eve's error probaility worse than Bob's one.

\section{Construction of CPPM}
 In this section, we discuss a method for the construction of  CPPM quantum signals  from PPM ones by the unitary operator associated with a symplectic transformation.

\subsection{Quantum Gaussian States}
            %
%
The classical probability distribution $\pi$ is characterized by 
the {\it characteristic function} $\phi(z)=\int \exp[ix^Tz]\pi(dx)$.
The characteristic function of Gaussian distribution 
with the mean $m$ and the correlation matrix $B$ is given as
$\phi(z)=\exp[im^Tz-\frac{1}{2}z^TB z]$.
We extend this  argument to  define the quantum Gaussian state \cite{Holevo:82}.
            We consider self adjoint operators on a Hilbert space ${\cal H}$,
$q_1,p_1,q_2,p_2,...,q_N,p_N$ satisfying Heisenberg CCR:
\begin{equation}\label{Heisenberg CCR}
\lbrack q_{j},p_{k}]=i\delta _{jk}\hbar
I,\;\;[q_{j},q_{k}]=0,\;\;[p_{j},p_{k}]=0,
\end{equation}
where 
$\delta_{j,k}=1$ for $j=k$ and $\delta_{j,k}=0$ for $j\neq k$.
            Let us introduce unitary operators 
\begin{equation}
V(z)=
\exp (\,i\,R^{T}z)
\end{equation}
for a real column $2r$-vector  
$z$  and 
$$
R=[q_{1},p_{1};\dots;q_{N},p_{N}]^{T}.
$$
The operators $V(z)$ satisfy the Weyl-Segal CCR
\begin{equation}
V(z)V(z^{\prime })=\exp \left[\frac{i}{2}\Delta (z,z^{\prime })\right ]V(z+z^{\prime }),
\label{weyl}
\end{equation}
where
\begin{equation}
\Delta (z,z^{\prime })=-z^T\Delta_N z^{\prime}
\end{equation}
is the canonical symplectic form with
\begin{equation} \label{Delta N}
\Delta_N=\bigoplus_{k=1}^{N}
\left [
\begin{array}{cc}
0&\hbar\\
-\hbar&0
\end{array}
\right ].
\end{equation}
            The Weyl-Segal CCR is the rigorous counterpart of the Heisenberg CCR, 
involving only bounded operators.
Now we can define the quantum characteristic function as
\begin{equation}
\tilde{\phi}(z)={\rm Tr} \rho V(z).
\end{equation}
            The transformation ${\cal L}$ satisfying \begin{equation}\label{S_cond}
\Delta({\cal  L}^Tz,{\cal L}^Tz^{\prime})=\Delta(z,z^{\prime})
\end{equation}
is called a {\it symplectic transformation}. 
We denote the totality of symplectic transformations by $Sp(N,\mathbb{R})$.
Eq. (\ref{S_cond}) can be rewritten as 
\begin{equation} \label{S_cond2}
{\cal L}\Delta_N {\cal L}^T=\Delta_N.
\end{equation}
The symplectic transformation preserves Weyl-Seagl CCR (\ref{weyl})
and hence it follows from Stone-von Neumann theorem 
that there exists the unitary operator $U$ satisfying [11]
\begin{equation}\label{stone von Neumann}
V({\cal L}^T z)=U^{\dagger}V(z)U.
\end{equation}
            We call such derived operator $U$ the {\it unitary operator associated with symplectic transformation} ${\cal L}$.
            %

The density operator $\rho$ is called Gaussian if its quantum characteristic
function has the form  
\begin{equation}
\tilde{\phi}(z)={\rm Tr}\rho V(z) =\exp\left[im^Tz-\frac{1}{2}z^TA z \right].
\end{equation}
In an $N$-mode Gaussian state,
$m$ is a $2N$-dimensional mean vector and $A$ is a
$2N\times 2N$-corralation matrix.
The mean $m$ can be arbitrary vector;
the necessary and sufficient condition on the correlation matrix $A$
is  given by
\begin{equation}\label{Robertson}
\Delta_N^{-1}A\Delta_N^{-1}+\frac{1}{4}A^{-1}\leq 0.
\end{equation}
            %

The coherent state $|\alpha \rangle$ ($\alpha=x+iy$) is the quantum Gaussian state
with the mean
\begin{equation}
m=\sqrt{2\hbar} ( x, y)^T 
\end{equation}
            and the correlation matrix
\begin{equation}\label{1mode squeezed}
A_1 =\frac{\hbar}{2}\left[ 
\begin{array}{cc}
1 & 0 \\ 
0 & 1
\end{array}
\right ],
\end{equation}
            and the $N$-ary coherent state $|\alpha_1\rangle\otimes \cdots \otimes |\alpha_N\rangle$
 ($\alpha_j=x_j+iy_j$)
is the quantum Gaussian state with the mean
\begin{equation} \label{C ave}
m=\sqrt{2\hbar} (x_1,y_1,.....,x_N,y_N  )^T
\end{equation}
            and the correlation matrix
\begin{equation}\label{C cor}
A_N=\bigoplus_{k=1}^{N}
\frac{\hbar}{2}
\left [
\begin{array}{cc}
1&0\\
0&1
\end{array}
\right ].
\end{equation}

\subsection{Generation of CPPM quantum signals 
by symplectic transformation}
            We study a way to generate CPPM quantum signals by the unitary operator  $U$ associated with a
symplectic transformation.
            Any unitary operator composed of beam splitters and phase shifts
can be described by a symplectic transformation. 
            First,  let us  consider the state $U|\phi\rangle$ for a general $N$-ary coherent state $|\phi\rangle=|\alpha_1\rangle\otimes \cdots \otimes |\alpha_N\rangle$.
By using Eq. (\ref{stone von Neumann}), the quantum characteristic function of the state $U|\phi\rangle$ is given as
\begin{equation}\label {CPPM ch}
\begin{split}
\tilde{\phi}(z)&={\rm Tr} U|\phi\rangle\langle\phi | U^{\dagger}V(z)
={\rm Tr} |\phi\rangle\langle\phi |  V({\cal L}^Tz)\\
&=\exp\left [ ({\cal L }m)^Tz-\frac{1}{2}z^T {\cal L}A_N{\cal L}^T z  \right ],
\end{split}
\end{equation}
where $m$ and $A_N$ is the mean vector and the correlation matrix 
given by  Eqs. (\ref{C ave}) and (\ref{C cor}) respectively.
            Eq. (\ref{CPPM ch}) shows that the state $U|\phi\rangle$ is the quantum
Gaussian state with the mean ${\cal L}m$ and the correlation matrix $ {\cal L}A_N{\cal L}^T$.
            Our interest is devoted to the case where the state $U|\phi\rangle$ is an $N$-ary coherent state.
Then the symplectic transformation $\cal{L}$ should satisfy
the condition ${\cal L}A_N{\cal L}^T=A_N$, that is,
\begin{equation}\label{S_cond4}
{\cal L}\in O(2N):=\{{\cal  L}\in M(2N,\mathbb{R}) | {\cal L}{\cal L}^T=I_{2N} \}
\end{equation}
where 
$I_{2N}$ is the $2N\times 2N$ identity matrix and
$M(2N,\mathbb{R})$ is the set of $2N\times 2N$ real matrices.
            %
Denoting the totality of $N\times N$-unitary matrices by $U(N)$,
we have the relation 
\begin{equation}
Sp(N,\mathbb{R})\cap O(2N) \cong U(N).
\end{equation} 
Here the matrix 
\begin{equation}\label{bs symp}
\begin{split}
{\cal L}=&
\begin{pmatrix}
r_{11}R(\theta_{11}) & \cdots & r_{1N}R(\theta_{1N}) \\
\vdots  &        & \vdots  \\
r_{N1}R(\theta_{N1}) & \cdots & r_{NN}R(\theta_{NN}) 
\end{pmatrix} \\
 & \in Sp(N,\mathbb{R})\cap O(2N),
\end{split}
\end{equation}
 with real numbers $r_{jk}$ and rotation matrices $R(\theta_{jk})$,
corresponds  to  the matrix
            \begin{equation}\label{bs C symp}
\begin{split}
{\cal L}_C=&
\begin{pmatrix}
r_{11}e^{i\theta_{11}} & \cdots & r_{1N}e^{i\theta_{1N}} \\
\vdots  &        & \vdots  \\
r_{N1}e^{i\theta_{N1}} & \cdots & r_{NN}e^{i \theta_{NN}} 
\end{pmatrix}\\
&\in U(N).
\end{split}
\end{equation}
            %
We can find that  the unitary operator associated with
${\cal L}_C\in U(N)$ transforms the state  $|\phi\rangle=|\alpha_1\rangle\otimes \cdots \otimes |\alpha_N\rangle$ to the state
$|\phi^{\prime}\rangle=|\alpha_1^{\prime}\rangle\otimes \cdots \otimes |\alpha_N^{\prime}\rangle$, 
where $\vec{\alpha}=
(\alpha_1,...,\alpha_N)
^T$ and $\vec{\alpha^{\prime}}=(\alpha_1^{\prime},...,\alpha_N^{\prime})^T$
are related in the equation
\begin{equation} \label{rel alpha}
\vec{\alpha^{\prime}}={\cal L}_C\vec{\alpha}.
\end{equation}
            In particular, from the PPM quantum signals $|\Phi_j\rangle=|0\rangle_1\otimes \cdots\otimes|\alpha_0\rangle_j\otimes\cdots \otimes|0\rangle_N$,  $j=1,2,..,N$, the CPPM ones are generated as  
\begin{equation} |\Psi_j\rangle=\otimes_{k=1}^N|\alpha_0r_{kj}e^{i\theta_{kj}}\rangle_k, 
j=1,2,...,N. 
\end{equation} 
In other words,  $N$-ary coherent states 
\begin{equation}
\otimes_{k=1}^N|\alpha_{kj}\rangle_k,j=1,...N,
\end{equation}
are the CPPM quantum signals  generated by applying the unitary operator
associated with ${\cal L}_C\in U(N)$ to the PPM quantum signals
$|\Phi_j\rangle$ if and only if 
the matrix with $(k,j)$-elements $\alpha_{k,j}/\alpha_0$ is unitary.

\section{Security Analysis of CPPM cryptosystem}

\subsection{ Heterodyne attack }\label{7_1}

We  give a foundation for discussing security of CPPM cryptosystem.
Without knowing the secret key $K_s$  Eve cannot apply  the appropriate unitary operator
to CPPM quantum signals, and hence she has to receive directly CPPM quantum signals. Since the quantum optimum receiver is unknown for such signals, we apply the heterodyne receiver,
which is suboptimum and appropriate to discuss the performance of error. This scheme is called heterodyne attack.  

Our main target in this subsection is to study  the  heterodyne attack on $U|\phi\rangle$, where $U$ is the the unitary operator associated with ${\cal L}_C\in U(N)$,
and $|\phi\rangle$ is a general $N$-ary coherent state  $|\alpha_1\rangle\otimes \cdots \otimes |\alpha_N\rangle$.

            %

Heterodyne detection is characterized by a family of operators with a parameter $\beta\in \mathbb{C}$,
\begin{equation}
X(\beta)=\frac{1}{\pi}|\beta\rangle\langle \beta |
\end{equation}
The outcomes $\beta$ of the heterodyne detection for a coherent state
$|\alpha\rangle$ appears with the probability density function
\begin{equation}
{\rm Tr}|\alpha\rangle\langle \alpha | X(\beta)=\frac{1}{\pi}|\langle \alpha | \beta \rangle |^2
=\frac{1}{\pi}e^{-|\alpha-\beta|^2},
\end{equation}
which represents the normal distribution with the correlation matrix $(1/2)I_2$.
            %

The outcomes $\vec{\beta}=(\beta_1,...,\beta_N)^T$ of the indivisual heterodyne detection 
for $U|\phi\rangle$
obeys the probability density function,
\begin{equation}\label{PDF}
\begin{split}
P_{U|\phi\rangle}(\vec{\beta})&={\rm Tr}U |\phi\rangle\langle \phi | U^\dagger \otimes_{j=1}^NX(\beta_j)\\
&={\rm Tr}U |\phi\rangle\langle \phi | U^\dagger\frac{|\psi\rangle\langle \psi|}{\pi^N}\\
&={\rm Tr} |\phi\rangle\langle \phi | \frac{U^\dagger|\psi\rangle\langle \psi|U}{\pi^N},
\end{split}
\end{equation}
            with
$|\psi\rangle=|\beta_1\rangle\otimes\cdots\otimes|\beta_N\rangle$.
            Here, putting $\vec{\beta^\prime}=(\beta_1^\prime,...,\beta_N^\prime)^T={\cal L}_C^*\vec{\beta}$
            and taking account of Eq (\ref{rel alpha}),
we get
\begin{equation}
\frac{U^\dagger|\psi\rangle\langle \psi| U}{\pi^N}=\otimes_{j=1}^NX(\beta^{\prime}_j).
\end{equation}
Note that $*$ represents the conjugate transpose and ${\cal L}_C^*$ corresponds to
the unitary operator $U^\dagger$.
            Substituting this equation to Eq. (\ref{PDF}), we obtain
\begin{equation}\label{PU}
P_{U|\phi\rangle}(\vec{\beta})=P_{|\phi\rangle}(\vec{\beta^{\prime}}),
\end{equation}
where $P_{|\phi\rangle}$ is the probability density function with which the outcomes of 
heterodyne detection for the state $|\phi\rangle$  appears.
            Eq. (\ref{PU}) shows that
the vectors $\vec{\beta^\prime}$ given by applying the unitary matrix ${\cal L}_C^*$ 
to the outcomes $\vec{\beta}$ obeys the probability density function $P_{|\phi\rangle}$.

\subsection{Error probability for CPPM quantum signal with key after measurement}

\begin{figure}[h!]
     \centerline{\includegraphics[width=\columnwidth]{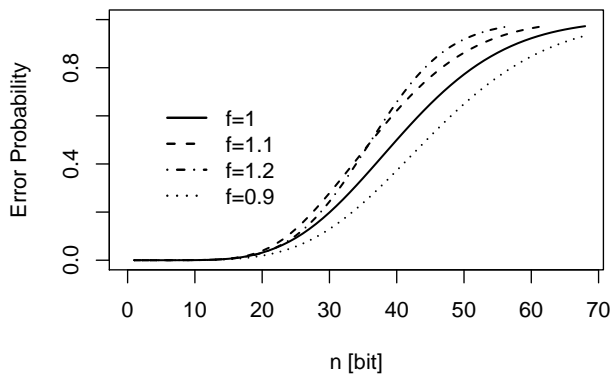}}
   \caption{Lower bounds for Eve's decoding error probability in the case where she gets the secret key $K_s$ after her  measurement for CPPM 
quantum signals}
   \label{fig1}
\end{figure}

It is difficult to evaluate the error performance for CPPM quantum signals by heterodyne attack, because the randomness of PRNG has to be taken into account.
Yuen showed the lower bound of the error performance by using heterodyne detection for the original PPM quantum signals [6]. But it may be not tight one. Here we try another approach.
We allow Eve to get the secret key $K_s$  after her measurement by heterodyne for CPPM quantum signals
and hence to know the unitary operator  $U_{K_i}$ and the corresponding unitary matrix ${\cal L}_{C,K_i}$.
Then, from  the discussions in the subsection \ref{7_1}, 
Eve can apply the unitary matrix ${\cal L}^*_{C,K_i}$ to
obtain the vector $\vec{\beta^{\prime}}$, which obeys 
to the probability density function $P_{|\Phi_j\rangle}$. 
This fact enables us to apply the decoding process for PPM signals. 
That is, Eve may use maximum-likelihood decoding for $\vec{\beta^{\prime}}$, 
whose rule is to pick the $j$ for which 
$\beta_j^{\prime}$ is largest, and her error probability is given as follows  [12]:
            \begin{equation}
P_e^{het}(key)=\int _{-\infty}^{\infty} \frac{1}{\sqrt{2\pi}}\exp\left[ -\frac{(y-\sqrt{2S})^2}{2}\right]Q_N(y)dy,
\end{equation}
where $S=|\alpha_0|^2$, and 
\begin{equation}
\begin{split}
Q_N(y)&=1-[\Phi(y)]^{N-1},\\
\Phi(y)&=\frac{1}{\sqrt{2\pi}}\int _{-\infty}^{y} \exp(-v^2/2)dv.
\end{split}
\end{equation}
 We will compute the lower bounds of Eve's error probability 
$P_{e}^{het}(key)$ to evaluate its convergence speed. 
The error probability $P_e^{het}(key)$ is lower bounded 
as \cite{Gallager}:
\begin{equation}\label{Kagen}
\begin{split}
P_e^{het}(key)\geq & \frac{1}{\sqrt{2\pi}}\int _{-\infty}^{z} \exp\left[ -\frac{(y-\sqrt{2S})^2}{2}\right]Q_N(y)dy\\
\geq & Q_N(z)\Phi(z-\sqrt{2S}),
\end{split}
\end{equation}
where the parameter $z$ can take any real number value.
            Putting $z=\sqrt{fn}$ and $n=\log _2 N$ in (\ref{Kagen}),
we obtain
\begin{equation}\label{Kagen limit}
P_e^{het}(key)\geq Q_{2^n}(\sqrt{fn})\Phi(\sqrt{fn}-\sqrt{2S})\to 1,n\to \infty.
\end{equation}
            Let us consider the case of $S=20$.
            Then Bob's error probability $P_e^{dir}$ is less than 
$10^{-8.69}$,
            while $P_e^{het}(key)$ converges to $1$.
Figure \ref{fig1} shows convergence behavior of lower bound
for $P_e^{het}(key)$.  In this figure, the lower bounds (\ref{Kagen})
for $f=0.9,1.1,1.2$,
are plotted with respect to $n=\log M$.
            Since the parameter $f$ in the lower bound (\ref{Kagen}) 
can take arbitrary real number, values of error probability $P_e^{het}(key)$
exist the region above the graphs in Figure \ref{fig1}.
Note that the above values of $f$ are chosen as they give better lower bounds for $P_e^{het}(key)$. 
            From Figure $\ref{fig1}$, it is found that the convergence speed of
lower bound for $P_e^{het}(key)$ is very slow;
$n>50$ ( $N > 2^{50} $ ) is needed to achieve the error probability 0.9.

Thus, in the CPPM scheme, Eve cannot pin down the information bit even if she gets the true secret key $K_s$ and PRNG
after her measurement, and consequently it has been proved that CPPM satisfies Eq(3).

\section{Subjects on system implementation}
According to the above analysis, one needs large number of $n$ when the signal energy is large.
Here we give a requirement of channel bandwidth for the secure communication by CPPM.
Let us assume that the signal band width is $W_s$ when there is no coding. 
In our scheme, first one has to transform the $n$ input information bit sequence to PPM signal with $2^n$ slots. Second, such PPM signals are converted into CPPM with the same number of slots.
If one wants to transmit such CPPM signal with no delay, the required bandwidth is 
\begin{equation}
W_{CPPM}=\frac{2^n}{n}W_s.
\end{equation}
Thus, the bandwidth  exponentially increases with respect to $n$. 
Since one needs the large $n$ to ensure the security, one needs a huge bandwidth.

On the other hand, we need to realize the unitary transformation to generate CPPM quantum signals from PPM ones. Such transformations may be implemented by combination of beam splitters and 
phase shifts [6], but to generate the CPPM quantum signals with uniform distance for all signal, we need also large number of elements. Thus we need more detailed consideration for the practical use. In future work, we will specify the realization method.

\section{Conclusion}
A crucial element of the coherent pulse position modulation cryptosystem is a generation of CPPM quantum signals from PPM ones
by a unitary operator.
In this paper, we have given a proper theory for a unitary operator and  a symplectic transformation basing on  
the quantum characteristic function.
Furthermore, by using the above results, we have shown the lower bound of error probability in the case where Eve gets the secret key after her measurement. This result clearly guarantees the fresh key generation supported by the secret key encryption system.

\section{Acknowledgment}
We are grateful to Dr. Usuda and the research staff in Tamagawa University for the discussion on this subjects.


\end{document}